\documentclass[twocolumn,showpacs,preprintnumbers,amsmath,amssymb,floatfix]{revtex4}

\usepackage{graphicx}
\usepackage{dcolumn}
\usepackage{bm}
\usepackage[bookmarks=true,colorlinks=true,linktocpage=true,backref=true]{hyperref}

\begin{document}

\preprint{Preprint}

\title{Stochastic quintessence models: jerk and fine-structure \\ variability constraints}

\author{Christine C. Dantas}

\affiliation{Instituto de Aeron\'autica e Espa\c co (IAE), \\ Departamento de Ci\^encia e Tecnologia Aeroespacial (DCTA), S\~ao Jos\'e dos Campos, 12.228-904, SP, Brazil}
\email{christineccd@iae.cta.br}

\author{Andr\'e L. B. Ribeiro}

\affiliation{Departamento de Ci\^encias Exatas e
Tecnol\'ogicas, Universidade Estadual
de Santa Cruz, Ilh\'eus, 45650-000 BA, Brazil}
\email{albr@uesc.br}

\date{\today}

\begin{abstract}

We report on constraints to the cosmological jerk parameter ($j$) and to possible variability of the fine-structure constant ($\Delta \alpha/\alpha$) based on stochastic quintessence models of dark energy, discussed by Chongchitnan and Efstathiou (2007). We confirm the results by these authors in the sense that many viable solutions can be obtained, obeying current observational constraints in low redshifts. We add the observables $j$ and $\Delta \alpha/\alpha$ to this conclusion. However, we find peculiarities that may produce, in the nearby Universe, potential observational imprints in future cosmological data. We conclude, for redshifts $z \lesssim 3$, that: {\it (i) } $j(z)$ fluctuates due to the stochasticity of the models, reaching an amplitude of up to $5\%$ relatively to the $\Lambda$CDM model value ($j_{\Lambda {\rm CDM}}=1$); and {\it (ii)} by contrasting two distinct (``extreme'') types of solutions, variabilities in $\alpha(z)$, linked to a linear coupling ($\zeta$) between the dark energy and electromagnetic sectors, are weakly dependent on redshift, for couplings of the order $|\zeta| \sim 10 ^{-4}$, even for large variations in the equation of state parameter at relatively low redshifts. Nonlinear couplings produce an earlier and steeper onset of the evolution in $\Delta \alpha/\alpha (z)$, but can still accommodate the data for weak enough couplings.

\end{abstract}

\maketitle


\section{Introduction \label{INTROD-SEC}}

Distance-redshift surveys of distant Type Ia supernovae point out that the universe
has recently entered a phase of accelerating expansion \cite{Rie98}, \cite{Per99}. The origin of this acceleration is not known and it is generally attributed to a dark energy component with negative pressure, inducing repulsive gravity and thus causing accelerated expansion \cite{Cop04}. The simplest candidate for this dark energy is the cosmological constant, $\Lambda$, which is expected to correspond to the zero-point
fluctuations of the quantum fields in QFT, giving rise to a vacuum energy density
$\rho_{vac}$ \cite{Sah00}. The cosmological constant enters as an additional source in the Einstein-Hilbert action, with no (or just weak) coupling to the other fields. Large scale observations put an upper bound on the $\Lambda$-term: $|\Lambda|<10^{-56}~{\rm cm^{-2}}$ or $|\rho_\Lambda | < 10^{-47}~{\rm GeV}^4$ \cite{Car92}. By contrast, theoretical estimates of various contributions to the vacuum energy density in QFT exceed this observational bound by many orders of magnitude \cite{Dol07}.

In the ${\rm \Lambda}$CDM model, the equation of state parameter is a constant, $w=-1$, while alternative dynamical models (e.g. quintessence, phantom field) this parameter is a function of time. Naturally, measurements of deviations from $w=-1$
would be decisive for a real breakthrough in cosmology. At the same time, as the Universe was once decelerating and is now accelerating, it is also useful to consider the third derivative of $a(t)$, the jerk parameter $j(t)$, to probe deviations from $w=-1$. An interesting way to use this parameter is the approach introduced by \cite{Bla04}. In that work, flat ${\rm \Lambda}$CDM models have a constant jerk with $j(a)=1$, thus, any deviation from $j=1$ measures a departure from ${\rm \Lambda}$CDM in the kinematical framework using a minimum of prior information, being independent of any particular gravity theory. The jerk parameter has been used to discriminate models of dark energy and modified gravity \cite{Sah03}, \cite{Ala07}, however, its observational status is presently quite poor. A critical discussion can be found in Ref. \cite{Boc13}, where it is argued that the value of the jerk parameter cannot be converged to high precision with current data. Yet, it is clear that a thorough investigation of the possible cosmological imprints of this observable is fundamental for discriminating dark energy models with future, higher-resolution experiment data. 

Another approach which have been increasingly recognized as a potential tool for constraining dynamical dark energy models comes from probing the uniformity and constancy of fundamental couplings \cite{Mar15b}, as for example the fine structure constant, $\alpha$. In this scenario, the dark energy and electromagnetic sectors are coupled through a dimensionless parameter, $\zeta$, which can be independently constrained. Recently \cite{Mar15}, a constraint at the two-sigma (95.4\%) confidence level has been obtained for this coupling, namely: $|\zeta | < 5 \times 10^{-6}$, assuming a dark energy with a constant equation of state. By relaxing this assumption, other constraints were found for a series of dynamical models \cite{Mar15b}, with bounds at the 2$\sigma$ level ranging from $|\zeta | < 5.6 \times 10^{-6}$ to  $|\zeta | < (0.8 \pm 1.2) \times 10^{-4}$, depending on the model.

In this report, we investigate the predictions for the jerk parameter and possible variations of the fine structure constant, both as a function of the redshift, $z$, based on the stochastically generated, dynamical models of dark energy proposed by Chongchitnan and Efstathiou \cite{Cho07} (hereon CE07). In this framework, viable models are selected out of solutions evolving from a quintessence-dominated regime at high z's, but which satisfy the observational constraints inferred at low z's ($z < 1$).  We specifically analyse whether $j(z)$, derived from these models, differs significatively from the pure $\Lambda$CDM model ($j_{\Lambda{\rm CDM}} \equiv 1$). We also obtain constraints to possible variations of $\alpha$ based on an ``extreme'' model (a ``late time'' viable model, as explained in detail in the next sections), parameterized by the coupling $\zeta$. 

This paper is organized as follows: in Sec. \ref{THEORY-SEC} we briefly review the Friedmann equations casted into energy phase-space variables, define the jerk parameter as a function of the scalar field, and the coupling between the dark and electromagnetic sections. We also summarize the method by CE07 for obtaining viable stochastic quintessence models. In Sec. \ref{RES-SEC}, we present a family of viable solutions and qualitatively analyse their behavior in the energy phase-space, classify the trajectories of selected models, and analyse the derived quantities of interest. In Sec. \ref{CONC-SEC}, we summarize our results.


\section{Theoretical framework for the stochastic models \label{THEORY-SEC}}

\subsection{The jerk parameter as a function of the scalar field \label{JERK-SEC}}

Starting from the Friedmann equations for a flat universe, namely,

\begin{equation}
\left({\dot{a}\over a}\right)={\kappa^2\over 3}(\rho_m + \rho_\phi), ~~
\left({\ddot{a}\over a}\right)= - {\kappa^2\over 6}(\rho_m + \rho_\phi + 3p_\phi), \label{Friedmann_Eqs}
\end{equation}

\noindent where $\kappa^2=m_{\rm Planck}^{-2} = 8\pi G$, and from the jerk parameter definition,

\begin{equation}
j(t)=+{1\over a}{d^3\over dt^3}\left[{1\over a}{da\over dt}\right]^{-3},
\end{equation}

\noindent one can find:

\begin{equation}
{dp_\phi\over dt}={2H^3\over 8\pi G}(1-j),
\end{equation}

\noindent where $H(t)=(\dot{a}/a)$. For a scalar field in the FRW background,

\begin{equation} 
\rho_\phi={1\over 2} \dot{\phi}^2 + V(\phi), p_\phi={1\over 2} \dot{\phi}^2 - V(\phi),
\end{equation}

\noindent where $V$ is the scalar field potential, and thus, in the slow roll approximation, we obtain:

\begin{equation}
j (\phi)\simeq 1 - {\kappa^2\over 2H^4}[V^\prime(\phi)]^2, \label{Jerk_Eq}
\end{equation}

\noindent where the prime indicates the derivative with respect to $\phi$.
Note that one can follow $j$ as a function of $\phi$ once
the potential is chosen, and $H\rightarrow H(\phi)$.  Also note that, by using the slow-roll approximation in Eq. \ref{Jerk_Eq}, the following constraint must be satisfied (e.g., \cite{Wei08}):

\begin{equation}
\epsilon_V \equiv {1 \over 16 \pi G} \left ( {V^{\prime} \over V} \right ) ^2 << 1. \label{EPS-V}
\end{equation}

\subsection{Brief review of the dynamics in the energy phase-space \label{DYN-SEC}}

The phase-space approach was introduced by Copeland, Liddle and Wands (1998), and here we only state the relevant equations for completeness and to fix our notation, which follows identically the presentation of CE07. The reader is referred to these papers for further details. We here also ignore any couplings between matter and the quintessence field, and the universe is assumed to be spatially flat as well. The canonical pair of variables in the energy phase-space are defined as:

\begin{equation}
x \equiv {\kappa \dot{\phi} \over \sqrt{6}H}, ~~ y \equiv {\kappa \sqrt{V} \over \sqrt{3} H}, \label{XY_Eq}
\end{equation}

\noindent with $x^2 + y^2 = 1 - \Omega_m -\Omega_r$, where $\Omega_m$ and $\Omega_r$ are the matter and radiation density parameters, respectively. 

Using the Friedmann equations (Eqs. \ref{Friedmann_Eqs}), one finds the following system of coupled differential equations in terms of the redshift ($z$):

\begin{eqnarray}
{dx \over dz} & = & -{1 \over 1 + z } \left [ -3x + \sqrt{3 \over 2} \lambda y^2 + 3 x^3+ {3 \over 2} x (1 +w_b ) (1 - x^2-y^2) \right ] \nonumber \\
{dy \over dz} & = & -{1 \over 1 + z } \left [ -\sqrt{3 \over 2} \lambda x y + 3 x^2 y + {3 \over 2}y(1+w_b)(1 - x^2-y^2)  \right ], \label{Phase_Eqs}
\end{eqnarray}
\noindent where $w_b$ is the matter-radiation equation of state parameter ($w_b = 0$ in the matter-dominated era, and $w_b = 1/3$ in the radiation-dominated era), and $\lambda$ is the ``roll'' parameter, defined as:

\begin{equation}
\lambda \equiv -{V^{\prime} \over \kappa V}. \label{Roll_Eq} 
\end{equation}

In these variables, the dark energy equation of state parameter is:

\begin{equation}
w_q  = {x^2 - y^2  \over x^2 + y^2},
\end{equation}

\noindent and the dark energy density parameter is:

\begin{equation}
\Omega_q  = x^2 + y^2.
\end{equation}

\subsection{The coupling between the electromagnetic and dark energy sectors \label{COUP-SEC}}

The stochastic quintessence model represents a dynamical scalar field scenario, and therefore a complete theoretical description requires the specification of how it couples to all other fields of the theory. We assume a coupling of the quintessence field with the electromagnetic sector only (e.g., \cite{Cal11},  \cite{Mar15}), given by the Lagrangean: 
\begin{equation}
\mathcal{L}_{\rm coup} = {1 \over 4}  - B_{\rm F} (\phi) F_{\mu \nu} F^{\mu \nu}, \label{EQ-LAG}
\end{equation}
\noindent where $F_{\mu \nu}$ is the electromagnetic field tensor, and $B_{\rm F} (\phi)$ is coupling function, assuming, in our present work \cite{Mar05}:
\begin{equation}
B_{\rm F} (\phi) = \left \{ 1 - \zeta_{\rm PC} [\Phi(z)]^q \right \} \exp [-\zeta_{\rm EC} \Phi(z)], \label{EQ-BF}
\end{equation}
\noindent with $q \in \mathbb{N}_+$, and the coupling parameters, $\{ \zeta_{\rm PC}, \zeta_{\rm EC}\} \in \mathbb{R}$ (dimensionless), are subject to independent observational constraints. The function $\Phi(z)$ is specified below (Eq. \ref{DA-FORM}). We then have the following possibilities:
\begin{itemize}
\item {Linear coupling (LC): $\{ q, \zeta_{\rm EC} \} =
\{1, 0 \}$; }
\item {Polynomial coupling (PC): $\{ q, \zeta_{\rm EC} \} =
\{\ge 2, 0 \}$; }
\item {Exponential coupling (EC): $\zeta_{\rm PC}=0$; }
\item {``Mixed'' coupling (MC): $\{\zeta_{\rm PC}, \zeta_{\rm EC} \} \neq 0$. }
\end{itemize}

To simplify notation, we will adopt $\zeta \equiv \zeta_{\rm PC}$ (in the LC and PC cases) or $\zeta \equiv \zeta_{\rm EC}$ (in the EC case), whenever no confusion may arise. In the MC case, we always specify each $\zeta_{\rm PC}, \zeta_{\rm EC}$.

Note that for small values of $\zeta_{EC}$ ($e^{-\zeta \Phi} \approx 1 - \zeta \Phi$), the EC becomes the LC-like case. Large values of $\zeta_{\rm EC}$ may violate observational constraints  \cite{Mar05}, so we will not consider them here. Therefore, we omit the purely EC case and consider the MC case with small values of $\zeta_{EC}$ only.

These couplings lead to a possible variation of the fine structure constant $\alpha$ as a function of cosmic evolution \cite{Cal11}, \cite{Mar15}:
\begin{equation}
{\Delta \alpha \over \alpha}(z) \equiv {\alpha(z) -\alpha(z=0) \over \alpha(z=0)} \equiv 1 - B_{\rm F}. \label{EQ-DAA}
\end{equation}
\noindent In terms of cosmological parameters, we have:
\begin{equation}
\Phi(z) \equiv \kappa (\phi -\phi_0)   =  \int_0^z \sqrt{3\Omega_q(z)(1+w_q(z))} {d z^{\prime} \over 1+ z^{\prime}}. \label{DA-FORM}
\end{equation}

\subsection{Viable models of quintessence from stochastic models \label{VIAB-SEC}}

In essence, the method by CE07 assigns stochastic values to the  ``roll'' parameter (Eq. \ref{Roll_Eq}), by uniformly binning the trajectory in the energy phase-space
(a solution to Eq. \ref{Phase_Eqs}), that is, $\lambda = \lambda(\Delta z)$, where $0 \leq \lambda \leq 10$, and $\lambda$ is a random variable. Therefore, in the process of integrating the coupled equations (Eqs. \ref{Phase_Eqs}), the stochastic values assigned for $\lambda$ are used in each step. The reader is referred to their paper for details of the binning method, which includes a bin refinement for low redshifts, as we strictly follow their procedure. 

The second important step in their method is to select only those solutions which are compatible with tight observational constraints, namely:

\begin{itemize}
\item{$-1 \leq w_q \leq -0.99$, at $z=z_{\rm obs}=0.3$;}
\item{$0.73 \leq \Omega_q \leq 0.74$, at $z=0$.}
\end{itemize}

As a third important step, we mention the derivation of the quintessence potential from
the energy variables. One integrates Eq. \ref{XY_Eq} in order to find the
quintessence field value as a function of z:

\begin{equation}
\phi(z) = -\sqrt{6} \int_0^z {x(\bar{z}) \over 1 + \bar{z}} d \bar{z},
\end{equation}

\noindent where the field variable is in units of $m_{\rm Planck}$ and is set to zero at $z=0$. Then, Eq. \ref{Roll_Eq} is integrated in order to find $V(z)$:

\begin{equation}
V(z) = V_0 \exp \left ( \sqrt{6}  \int_0^z {\lambda (\bar{z}) x(\bar{z}) \over 1 + \bar{z}} d \bar{z} \right ), \label{VZ_Eq}
\end{equation}

\noindent where $V_0$ is $\Omega_q (z=0)$. Hence, to find $V(\phi)$, one just combines the results of the two previous integrations.

\begin{figure} [tbhp]
\includegraphics[width=1\columnwidth]{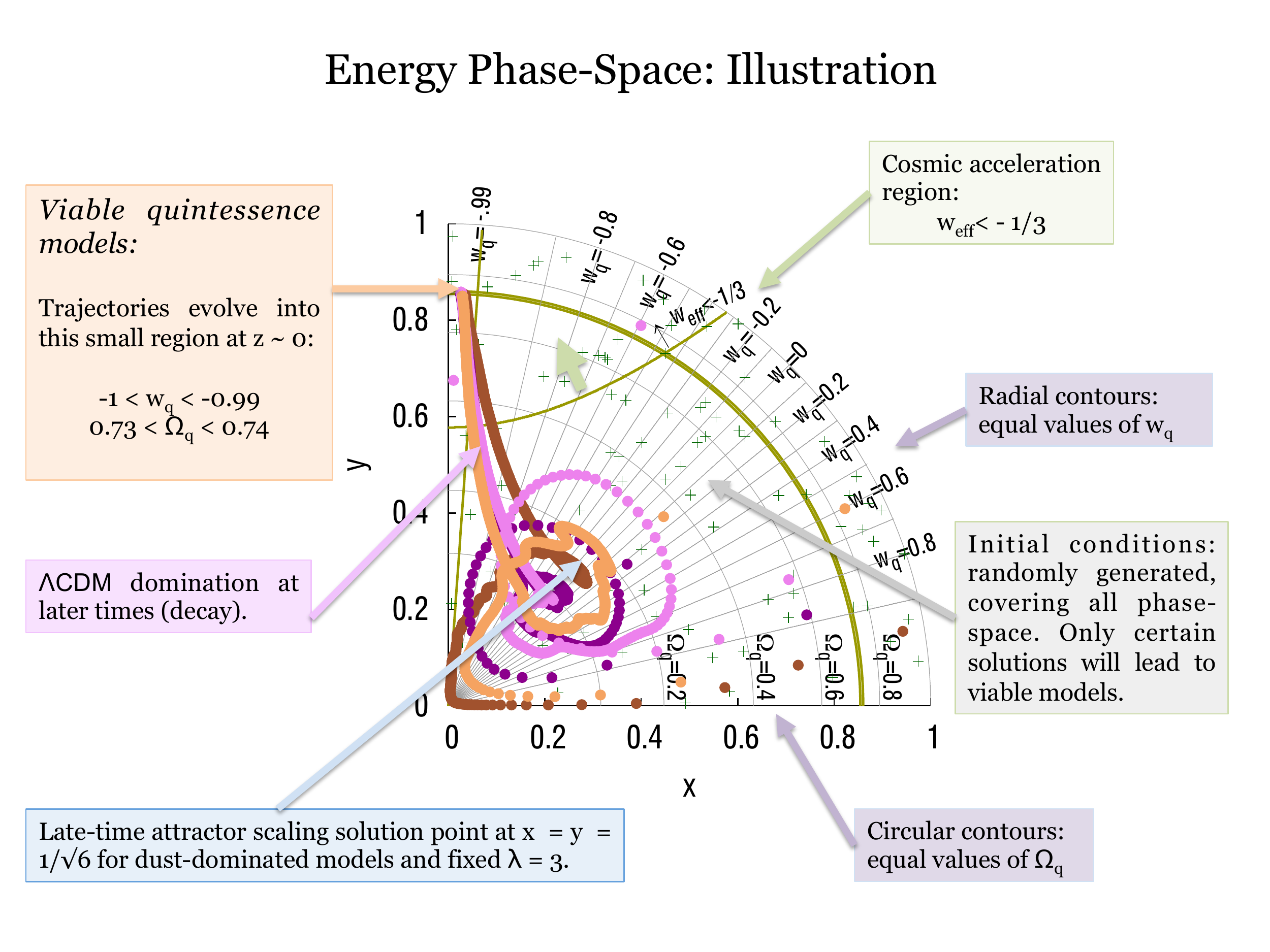}
\caption{(Color online). Illustration of the main features of the energy phase-space and general characteristics of the stochastic trajectories of the CE07 proposal. \label{Fig-Illustr}}
\end{figure}

In order to clarify how the dynamical variables should be read in the energy phase-space, according to the exposition of Sec. \ref{DYN-SEC}, we present in Fig. \ref{Fig-Illustr} an illustration indicating the main features of the energy phase-space and the general characteristics of the stochastic trajectories of the CE07 proposal. This illustration should serve as a guide to the actual obtained models, to be discussed in Sec. \ref{ENS-SEC}, since at first the various solution trajectories will be presented collectively in the same phase-space, which tend to produce an overloaded picture. The models will be subsequently analysed into different sub-classes.

\clearpage

\onecolumngrid

\twocolumngrid

\section{Models and Results \label{RES-SEC}}

\subsection{Ensembles of stochastic quintessence models \label{ENS-SEC}}

We consider dynamical models with solutions evolving from a quintessence-dominated regime in high redshifts and decaying towards a behavior that mimics the cosmological constant in low redshifts, in agreement with the observational constraints mentioned in the previous section. In order to produce these models, the ``roll'' parameter, $\lambda$ (c.f. Eq. \ref{Roll_Eq}), must sharply decay to sufficient small values in low redshifts. As explained in CE07 the corresponding potential presents a transition from a steep to a shallow slope at some characteristic field value. 

In order to produce the above-mentioned potential decay, we imposed amplitudes of up to $\lambda \leq 10$ in higher redshifts and up to $\lambda \leq 0.1$ in lower redshifts. The latter amplitude is somewhat arbitrary but must be small enough to produce compatible models. An inspection of Fig. 2 (second panel) of CE07 indicates that $\lambda \leq 0.1$ would be an adequate setting. Indeed, after preliminary tests looking for solutions around that value, viable models could be found for that choice, so it was subsequently fixed in our analysis.

Note, however, that the choice of the upper range of $\lambda$ in low redshifts has an important consequence for the resulting models, particularly for the jerk parameter.  The approximation given by Eq. \ref{Jerk_Eq} depends on the derivative of the potential ($V^{\prime}$). On the other hand,  Eq. \ref{VZ_Eq} states that the potential may fluctuate around some level, depending on fluctuations in $\lambda(z)$, which are convoluted with $x(z)$ in the integral. This fact may leave an imprint on the derivative $V^{\prime}$, and consequently, on the jerk parameter. This will be discussed in the next section.

\clearpage

\onecolumngrid

\begin{figure} [tbhp]
\includegraphics[width=1\columnwidth]{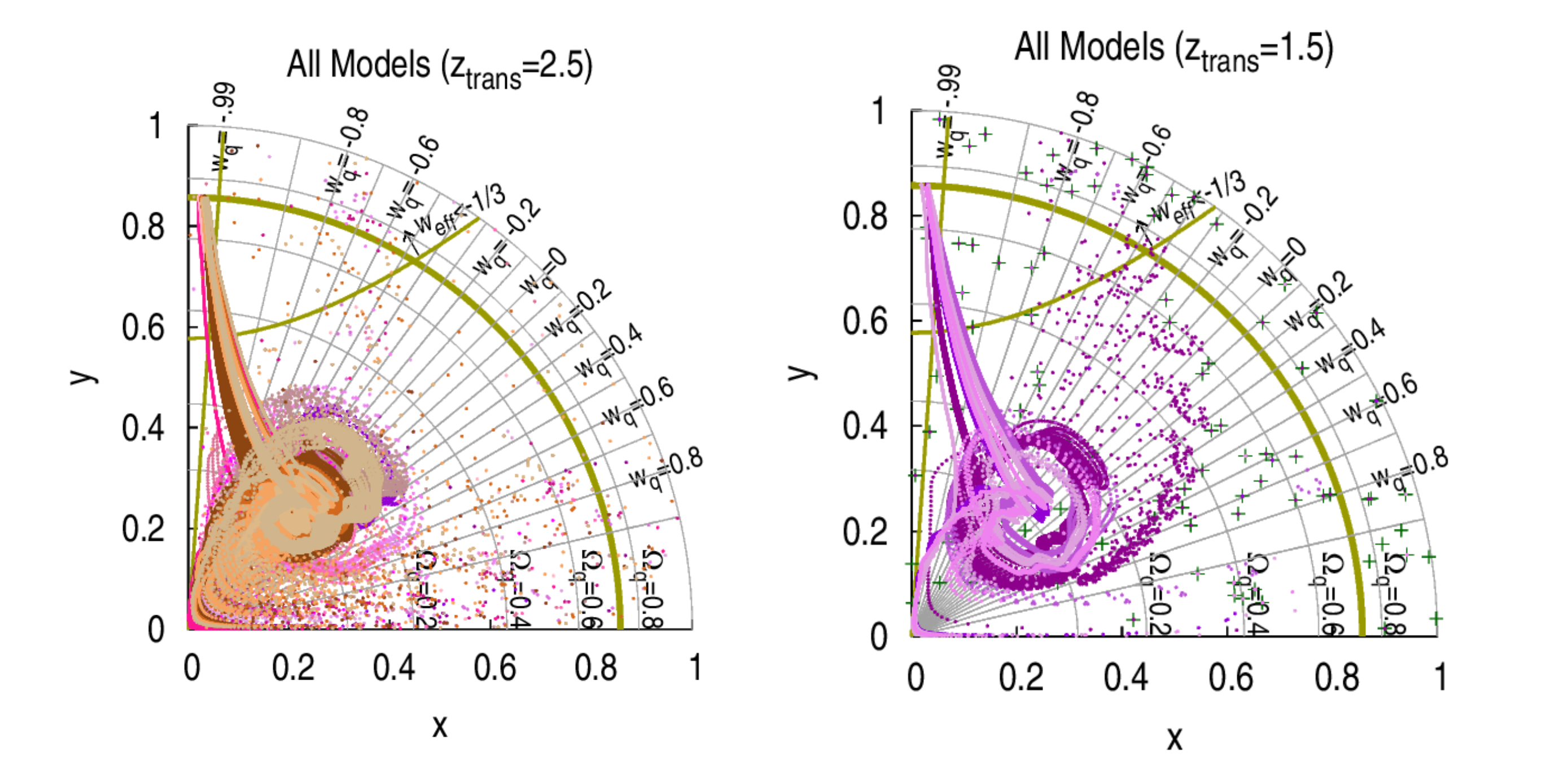}
\caption{(Color online). Trajectories of viable stochastic quintessence models in the energy phase-space. Left panel: the ($z_{\rm trans}=2.5$)-ensemble models, obtained from different sets of initial conditions and $\lambda(z)$ distributions (for each of the $20$ sets of the ensemble). Right panel: the ($z_{\rm trans}=1.5$)-ensemble models, obtained a fixed set of initial conditions but different $\lambda(z)$ distributions (for each of the $5$ sets of the ensemble). Different trajectory colors indicate different sets (or, equivalently, different $\lambda(z)$ distributions). The ``+'' symbols in the phase space of the ($z_{\rm trans}=1.5$)-ensemble models (right panel) indicate the (fixed) initial conditions taken from a set of the ($z_{\rm trans}=2.5$)-ensemble (see text). \label{Phase-ALL-COMB-prep}}
\end{figure}
\twocolumngrid

We generated two different ``ensembles'' of solutions, according to two high/low-$\lambda$'s transition redshifts, namely: $z_{\rm trans}=\{1.5, 2.5\}$. These values basically cover the transitions used in CE07 (see their Fig. 2, second panel). We considered the ($z_{\rm trans}=2.5$)-ensemble as the ``fiducial'' one, and the ($z_{\rm trans}=1.5$)-ensemble was used for comparison purposes, as we explain later. These ensembles differ not only on the choice of $z_{\rm trans}$, so we discuss them briefly:


\begin{itemize}
\item {For the ($z_{\rm trans}=2.5$)-ensemble, $20$ sets of solutions were produced, tentatively starting from $100$ random initial conditions $(x_0,y_0)$ on the phase-space, {\it generated anew for each set}. A stochastic $\lambda(z)$ distribution was also {\it generated anew for each set}. Therefore, each set contains a varying number of viable solutions (or trajectories in phase-space). This ensemble has a total number of $321$ solutions. }
\item {For the ($z_{\rm trans}=1.5$)-ensemble, $5$ sets of solutions were obtained, {\it but all of them starting from the same tentative initial conditions}. Otherwise, just as the ($z_{\rm trans}=2.5$)-ensemble, the $\lambda(z)$ distributions were also {\it generated anew} for each set. The initial conditions chosen  correspond to those used in a specific set of the ($z_{\rm trans}=2.5$)-ensemble. The reason for that choice is given Sec. \ref{TYPES-SEC}. Each set also produced a varying number of viable solutions. This ensemble has a total number of $101$ solutions.}
\end{itemize}

In summary, the sets are a label for a distinct $\lambda(z)$ distribution. The ($z_{\rm trans}=2.5$)-ensemble carries a new initial condition for each of its $20$ sets, while the ($z_{\rm trans}=1.5$)-ensemble uses a fixed set of initial conditions for its $5$ sets. Notice that a given set may have produced, say, $n$ viable trajectories, with $3\leq n \leq 100$ (we have discarded sets producing $n\leq 2$ solutions). This means that, out of the $100$ tentative initial conditions for that set, ($100-n$) ones were not ``successful'' in generating viable trajectories. Hence, each set happens to have a varying number of solutions, with no predictable pattern due to the nonlinear nature of the equations and the stochastic character of the $\lambda(z)$ distribution.
Evidently, while producing those sets, not every integration run could produce viable solutions for a given tentative set of initial conditions and $\lambda(z)$ distribution, so they were searched for by trior and error. The sets of solutions reported above are therefore the end result of that process.

In Fig. \ref{Phase-ALL-COMB-prep}, all the obtained solutions (trajectories) on the phase-space are shown for each of the two ensembles ($z_{\rm trans}=\{1.5, 2.5\}$). All these trajectories represent viable stochastic quintessence solutions obeying the observational constraints at low redshifts. The reader should compare those figures with Fig. 2 (top panel) of CE07, which show good agreement.

As pointed out in CE07, there are basically two main categories of models: (1) those obtained from flat potentials, like static ($\lambda(z) \sim 0, \forall z$) and ``skater'' models ($x\sim 0$), and (2) dynamical models (with trajectories evolving more broadly in phase-space), and both classes can be reproduced in stochastic quintessence models, leading to viable solutions. One should also note that these are locally (``time sliced'') exponential potential models (c.f. Eq. \ref{VZ_Eq}) and their behavior can be inferred from the attractor dynamics in energy phase-space for the purely exponential potential (see Ref. \cite{Cop98} for a complete analysis). Our results show (Fig. \ref{Phase-ALL-COMB-prep}) that the models move towards/around the known late-time attractor scaling solution at $x=y=\sqrt{1/6}$ (dust-dominated model with fixed $\lambda=3$, c.f. Ref. \cite{Cop98}), and subsequently decay towards the scalar field dominated solution in lower redshifts (after the given transition redshift; $z_{\rm trans}=\{1.5, 2.5\}$).

One should recall that the ($z_{\rm trans}=2.5$)-ensemble models use, for each set, a large variety of different initial conditions and $\lambda(z)$ distributions, whereas the ($z_{\rm trans}=1.5$)-ensemble models are all drawn from the same initial conditions, but with  different $\lambda(z)$ distributions (for each set). Therefore, the latter ensemble shows that a variety of trajectories are possible for the same initial conditions, with the dynamics being indeed dictated by the $\lambda(z)$s. At this level, it is difficult to infer significant differences in the behavior of the solutions between both ensembles. In order to advance this understanding, particularly with respect to potential observational imprints, we have selected solutions according to finer criteria.

\subsection{Early and Late Models \label{EL-SEC}}

 For each set (or alternatively, for each $\lambda(z)$ distribution), we asked how early/late a trajectory meets the observational constraint of $w_q=-0.99$, by selecting, in this sense,  the ``earliest'' and ``latest'' trajectories in each set. In addition we imposed a redshift cutoff restriction  (solutions were discarded if not meeting the criteria):

\begin{itemize}
\item{EARLY MODELS: the ``earliest'' solution for each set, reaching $w_q=-0.99$ {\it earlier than} $z_w = 0.55$.}
\item{LATE MODELS: the ``latest'' solutions for each set, reaching $w_q=-0.99$ {\it later than} $z_w = 0.45$.}
\end{itemize}

\begin{figure} [bhtp]
\includegraphics[width=1\columnwidth]{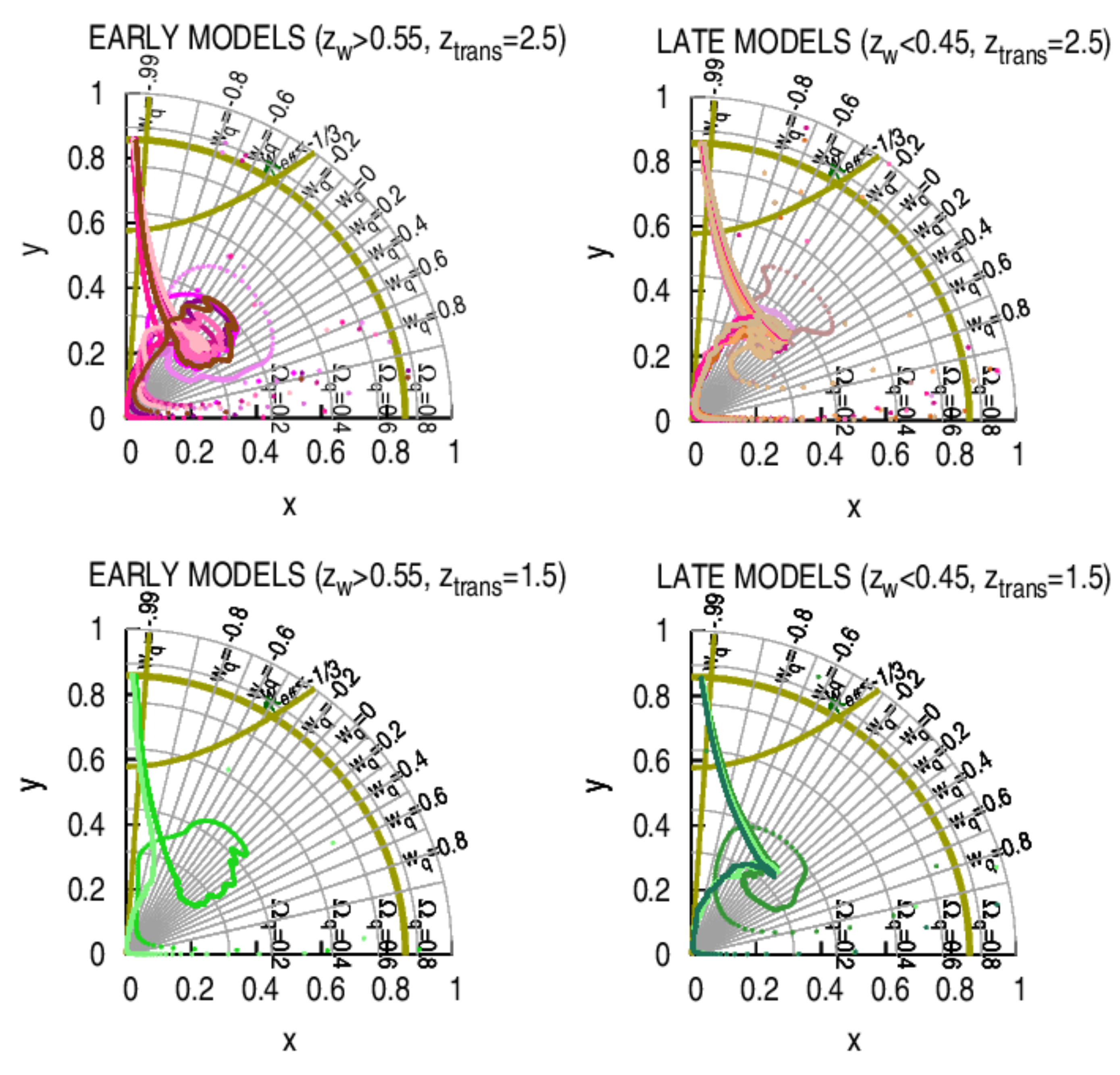}
\caption{(Color online). Trajectories of viable stochastic quintessence models in the energy phase-space, for the selected models: EARLY (left panels) vs. LATE (right panels), from the ($z_{\rm trans}=2.5$)-ensemble (top panels) and the ($z_{\rm trans}=1.5$)-ensemble (bottom panels).  \label{Phase-EARLY-LATE-COMB-prep}}
\end{figure}

Notice that most of the original models do not meet the above criteria and were therefore left out in the subsequent analysis [only $26$ models met the criteria, with $21$ from the ($z_{\rm trans}=2.5$)-ensemble]. In Fig. \ref{Phase-EARLY-LATE-COMB-prep}, we present the phase-space of the selected models. The trajectories shown in this figure form a subset of those in Fig. \ref{Phase-ALL-COMB-prep}. 

As mentioned previously, the attractor dynamics in energy phase-space for the fixed exponential potential has been thoroughly studied previously \cite{Cop98}, and it is not the main focus of the present investigation, but we mention two qualitative points: (1) the transition redshift does {\it not} seem to constrain the early/late classification, at least at the level here imposed, in the sense of, e.g., expecting EARLY models to be preferably found in ($z_{\rm trans}=2.5$)-ensemble models and/or LATE models to be preferably found in ($z_{\rm trans}=1.5$)-ensemble models. (2) These ``extreme'' models, however, seem to evolve preferentially (in higher redshifts) through the ``horizontal branch'' ($y^{\prime} \sim 0$), along paths satisfying $y<0.2$, before decaying towards the scalar field dominated solution. Since this characteristic seems to be a common feature of these ``extreme'' models, we focus on a qualitative analysis of their subsequent evolution in the next section.

\subsection{Types of Evolution \label{TYPES-SEC}}

In this section we identify main types of possible evolution paths in phase space from redshifts preceding the transition redshift. Given that the ($z_{\rm trans}=1.5$)-ensemble selections (EARLY and LATE models) are composed of only few representatives, they display more clearly the behavior mentioned in point (2) of the previous section, which we focus here. These models allow us to clearly identify different intermediary evolutions, {\it after moving along the ``horizontal branch''} ($y^{\prime} \sim 0$). We label these paths according to the subsequent evolution as the following types: 
\begin{itemize}
\item{{\it type-(i)}-- the path never exceeds $w_q \sim -0.7$.}
\item{{\it type-(ii)}--  the path circulates around a region nearby the late-time attractor scaling solution.}
\item{{\it type-(iii)}-- the path is ``peaked'' at some point nearby the late-time attractor scaling solution.}
\end{itemize}

\begin{figure} [tbhp]
\centering
\includegraphics[width=1\columnwidth, height=1.1\columnwidth]{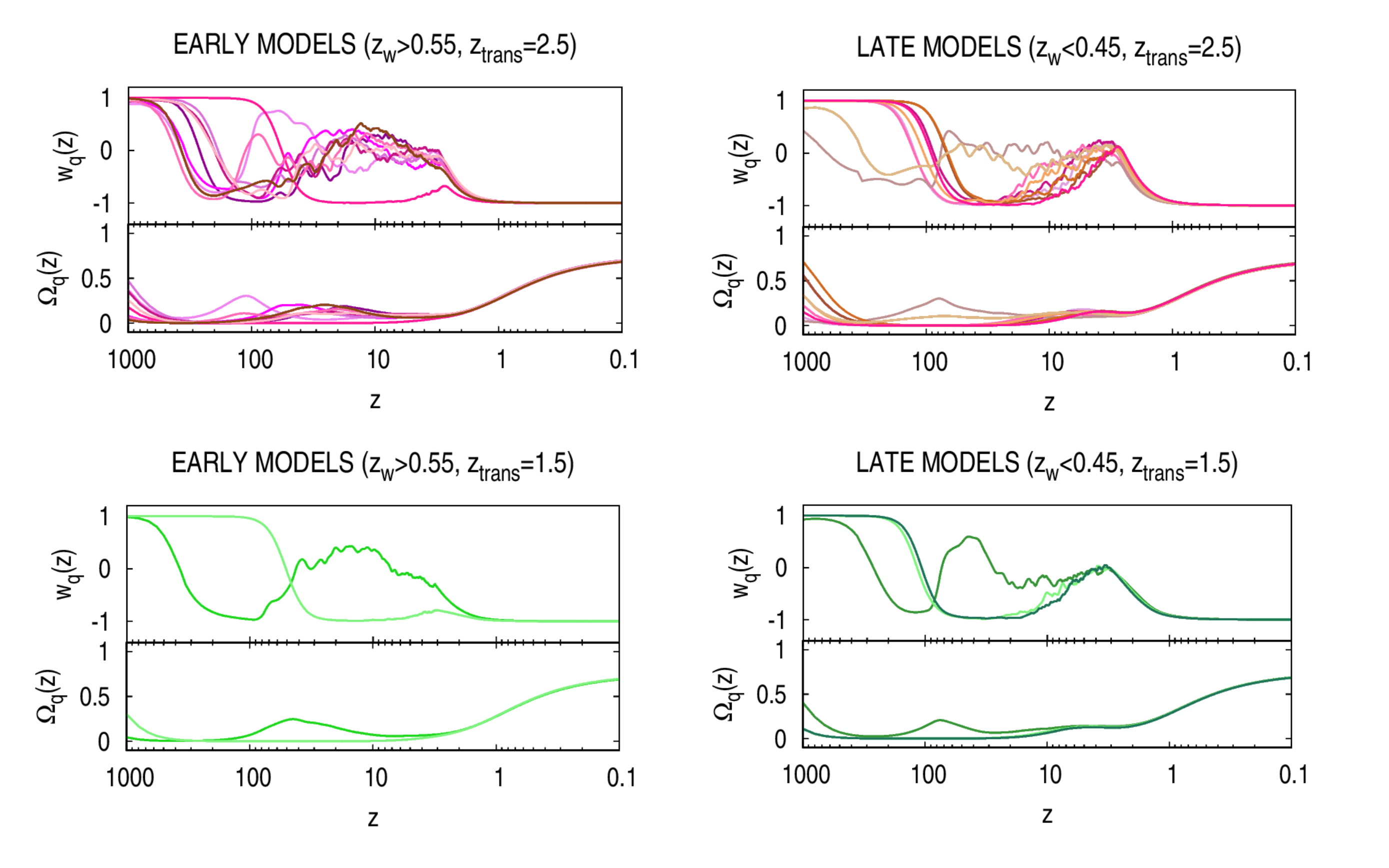}
\caption{(Color online). The behavior of cosmological parameters of the selected solutions as a function of redshift: the dark energy equation of state parameter $w_q(z)$ (top panel in each figure) and the dark energy density parameter $\Omega(z)$ (bottom panel in each figure). Figures are ordered as in Fig. \ref{Phase-EARLY-LATE-COMB-prep}:  EARLY (left panels) vs. LATE (right panels), from the ($z_{\rm trans}=2.5$)-ensemble (top figures) and the ($z_{\rm trans}=1.5$)-ensemble (bottom figures). \label{W-COMB}}
\end{figure}

\begin{figure} [hbtp]
\centering
\includegraphics[width=1\columnwidth]{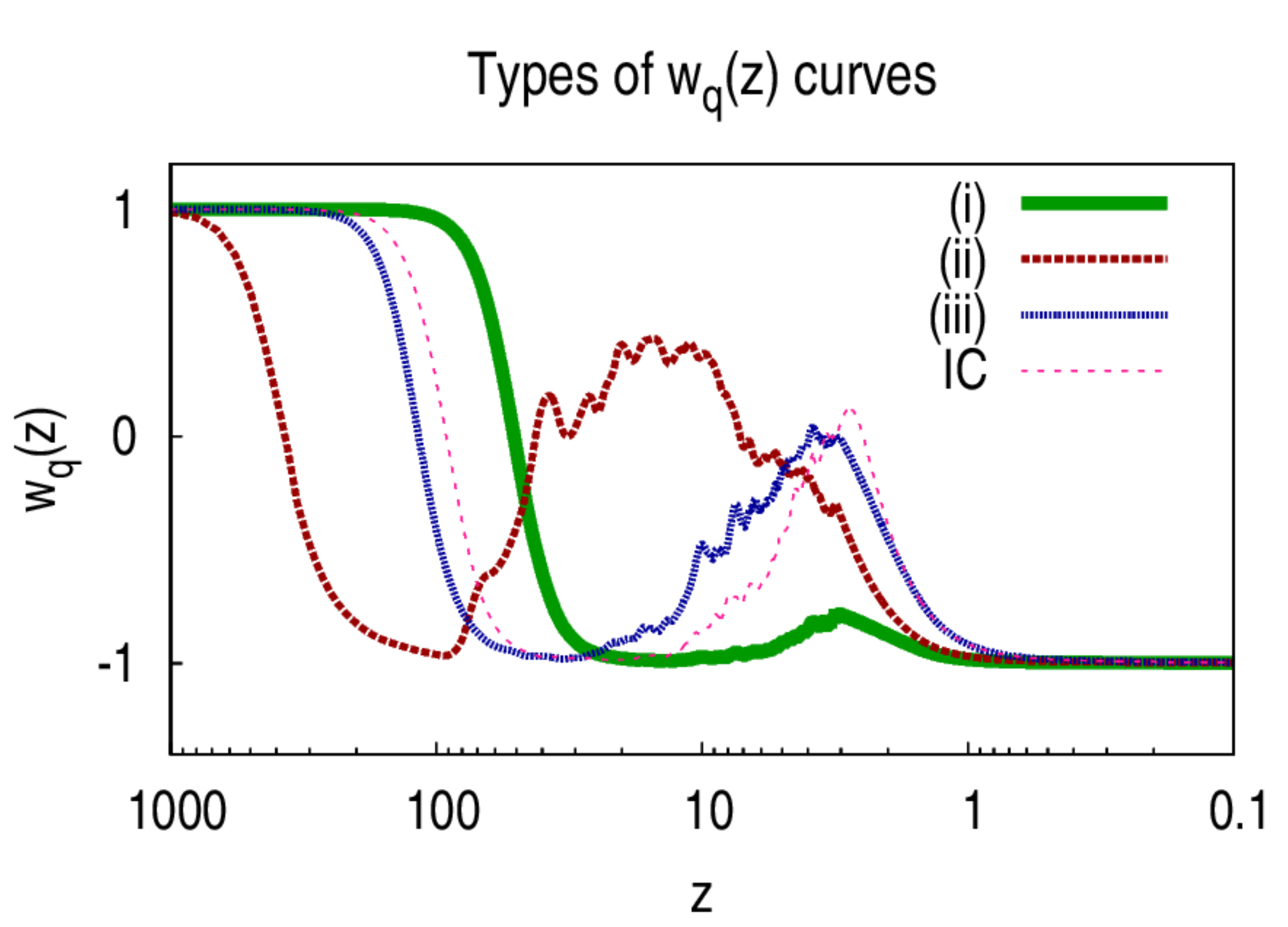}
\caption{(Color online).  Representative curves of type-(i)---(iii) behaviors in the evolution of the dark energy equation of state parameter $w_q(z)$. The ``latest'' model obtained from the ($z_{\rm trans}=2.5$)-ensemble is indicated by the ``IC'' curve, and its initial conditions were fixed for all the sets generated by the ($z_{\rm trans}=1.5$)-ensemble. \label{TYPES}}
\end{figure}

These types of paths can be clearly identified by an examination of another display, shown in Fig. \ref{W-COMB}, where  we present the behavior of cosmological parameters of the selected solutions as a function of redshift: the dark energy density parameter $\Omega_q(z)$ and the dark energy equation of state parameter $w_q(z)$ (compare them with the corresponding panels in Fig. 2 of CE07). The type-(i)--(iii) behaviors give distinct forms to $w_q(z)$, which are illustrated in Fig. \ref{TYPES}.

Type-(i) paths correspond to  a $w_q(z)$ with a ``step function'' behavior, with a little ``bump'' before the transition redshift. This agrees with the observation that these models do not ``move around'' the late-time attractor scaling solution, as types-(ii) and -(iii) do. Type-(ii) trajectories show a ``noisy'', ``periodic''-like behavior in $w_q(z)$ between redshifts $100$ and the transition redshift, corresponding to the ``circulatory'' behavior in phase-space. Type-(iii) solutions show a very similar behavior to the type-(i) ones at high redshifts, but show a larger, ``gaussian''-like profile before the transition redshift, corresponding to the ``peaked'' behavior of the path in phase-space. Type-(i) and -(ii) are favored in EARLY models, whereas type-(iii) is favored in LATE models, regardless of the transition redshift used, and all these properties are not sensitive to the initial conditions (c.f. Fig. \ref{Phase-ALL-COMB-prep}).

Indeed, given the stochastic evolution of the models, the lack of sensitivity on initial conditions is somewhat expected. This was tested by setting the same initial conditions for all sets in the ($z_{\rm trans}=1.5$)-ensemble. The initial conditions used were taken from a set of the ($z_{\rm trans}=2.5$)-ensemble, indicated in Fig. \ref{TYPES} as the ``IC'' curve. It was the ``latest'' model obtained in the latter ensemble, clearly seen as a type-(iii) solution. Yet, the ($z_{\rm trans}=1.5$)-ensemble models exhibit all types of curves for these same initial conditions.

It is interesting to note that type-(i) trajectories can mimic the cosmological constant at redshifts higher than $10$, but presents a small deviation from the purely $\Lambda$-CDM model at around the transition redshift, with a width of about $\Delta z \sim 3$. If observed, such a deviation could be a signature favoring a stochastic quintessence model of this type.

\subsection{The behavior of $j(z)$ \label{JERKDEZ-SEC}}

\begin{figure} [hbtp]
\includegraphics[width=1\columnwidth,height=1.02\columnwidth]{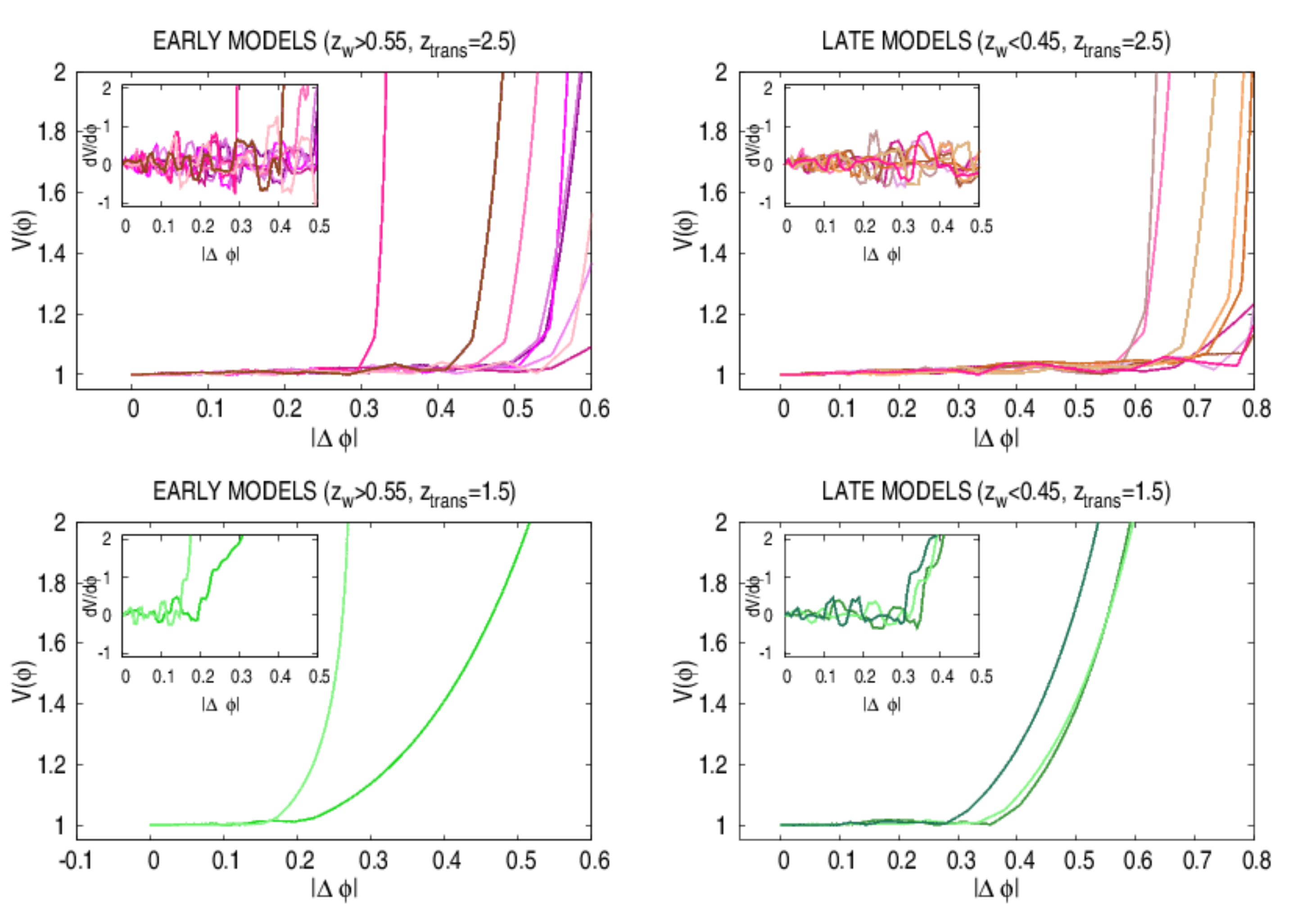}
\caption{(Color online). The potential of the models and the corresponding derivatives $dV/d\phi$ (insets). Figures are ordered as in Fig. \ref{Phase-EARLY-LATE-COMB-prep}:  EARLY (left panels) vs. LATE (right panels), from the ($z_{\rm trans}=2.5$)-ensemble (top figures) and the ($z_{\rm trans}=1.5$)-ensemble (bottom figures).} \label{VdePhi-INSET-COMB-prep}
\end{figure}

 In Fig. \ref{VdePhi-INSET-COMB-prep} we present the obtained potential $V(\phi)$ of the selected models as well as the corresponding derivatives $dV/d\phi$ (insets). The reader should compare those figures with Fig. 4 of CE07, which show similar, ``exponential-like'' behavior. Note also that these curves are in qualitative agreement with the condition given by Eq. \ref{EPS-V}, in the late-time regime (after $z_{\rm trans}$), so that the ``slow-roll'' approximation, given by Eq. \ref{Jerk_Eq}, can be used at very low redshifts.


\begin{figure} [hbtp]
\includegraphics[width=1.0\columnwidth,height=1.0\columnwidth]{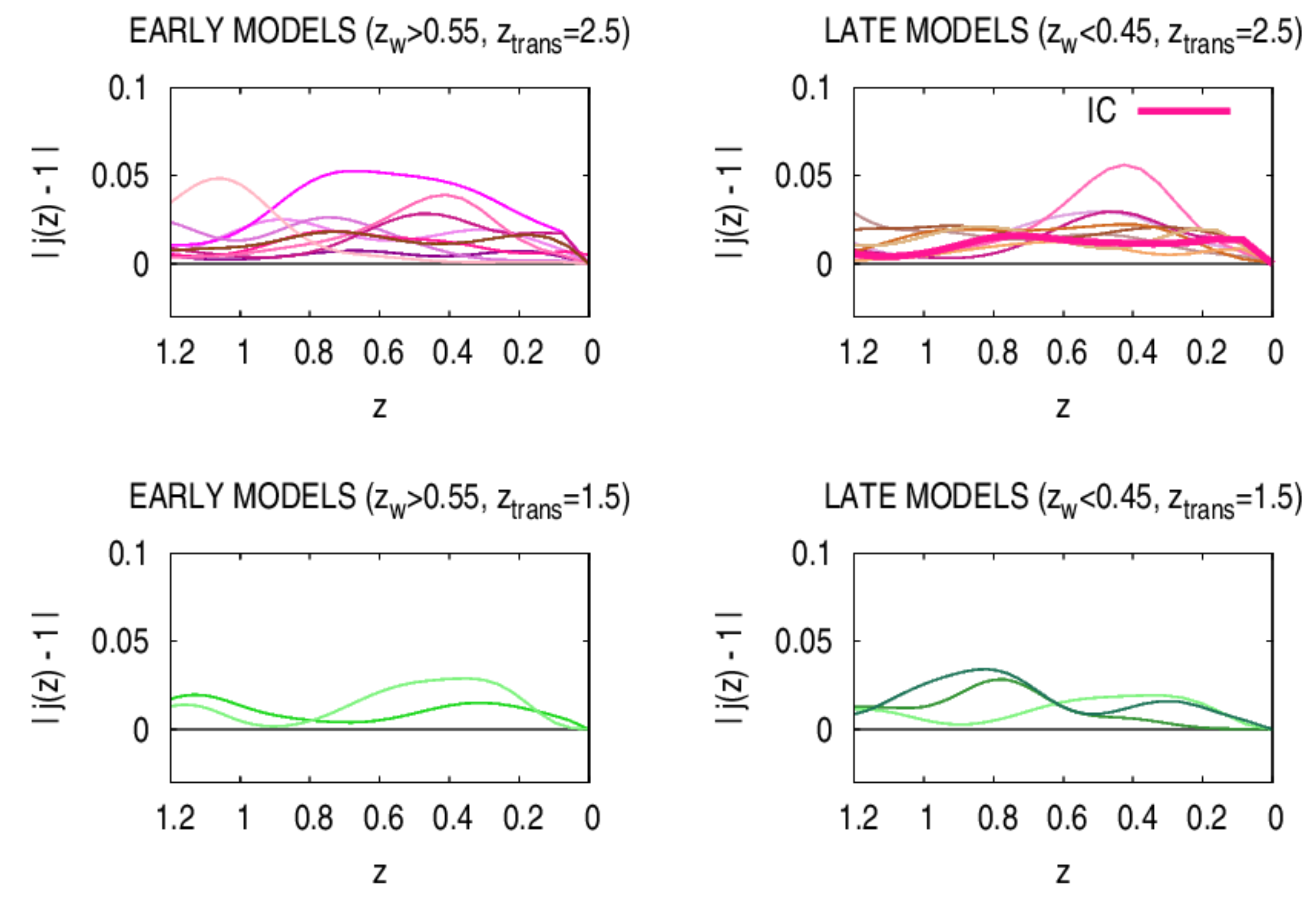}
\caption{(Color online). The jerk parameter of the models as compared to $\Lambda$CDM. The horizontal line at $|j(z)-1| = 0$ the  signals a $\Lambda$CDM behavior ($\equiv j_{\Lambda{\rm CDM}}$). Figures are ordered as in Fig. \ref{Phase-EARLY-LATE-COMB-prep}:  EARLY (left panels) vs. LATE (right panels), from the ($z_{\rm trans}=2.5$)-ensemble (top figures) and the ($z_{\rm trans}=1.5$)-ensemble (bottom figures). The ``latest model'' (labeled as ``IC'' in Fig. \ref{TYPES}) is indicated in the top right panel, as a thick line. \label{Jerk_VS-LCDM-COMB-prep}}
\end{figure}

The very steep parts of the potential curves are found for larger values of $|\Delta \phi |$ in the selected ($z_{\rm trans}=2.5$)-ensemble models, regardless of being EARLY or LATE models, with the exception of one EARLY model, showing the steep ascent starting at $|\Delta \phi | \sim 0.3$. This model is identified as a type-(i) path. The selected ($z_{\rm trans}=1.5$)-ensemble models show steep curves starting at smaller values of $|\Delta \phi |$, and they seem to ascend more gradually than the former models, except for an EARLY model, also identified as a type-(i) path, which ascends abruptly at $|\Delta \phi | \sim 0.2$.

It is clear from the insets of Fig. \ref{VdePhi-INSET-COMB-prep} that the derivatives of the potential are noisy (we disregard steep parts of the potential), and this can be traced back to the amplitude of the $\lambda(z)$ distribution after the transition redshift. Even though the amplitude of the later is small, the impact on $d V/ d\phi$ can be significant.

In order to obtain the behavior of the jerk parameter (c.f. Eq. \ref{Jerk_Eq}) for the ``extreme'' models defined in the previous section, we must calculate the derivatives of the potential, together with the Hubble parameter, $H(z)$, which is obtained directly from the solutions (c.f. Eq. \ref{XY_Eq}).

In Fig. \ref{Jerk_VS-LCDM-COMB-prep}, we present a comparison of the obtained jerk parameter of the selected models relatively to a pure $\Lambda$CDM model, namely the absolute deviations $|j(z)-1|$ as a function of redshift. Due to the nature of the derivatives of the potential, the derived noisy $j(z)$ distributions were smoothed out using B\'ezier curves linking the jerk solution points, for a better visualization. The $j(z)$ generally diverges before the transition redshift, a regime where Eq. \ref{EPS-V} is no longer valid. However, for lower redshifts the models offer an upper bound to potential observational fluctuations in $j(z)$. Our  results indicate that the jerk parameter may fluctuate up to $5\%$ from the pure $\Lambda$CDM model, and still comply with all the observational constraints in low redshifts.

\clearpage

\onecolumngrid

\begin{figure} [bthp]
\includegraphics[width=1\columnwidth]{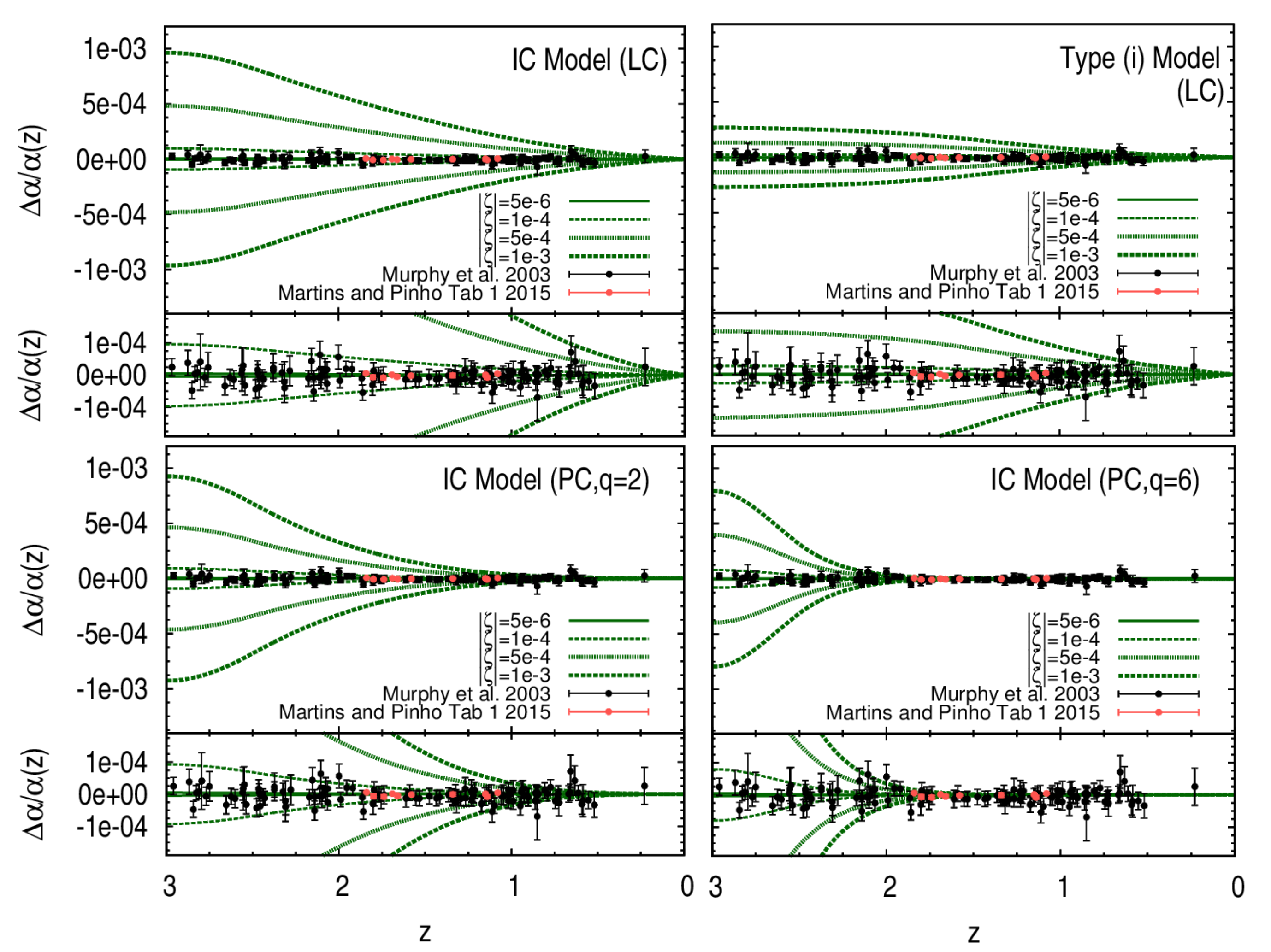}
\caption{(Color online). Stochastic quintessence predictions for the variation $\Delta \alpha/\alpha (z)$ for polynomial couplings (PC). Curves are parametrized by the coupling factor $|\zeta | \equiv |\zeta_{\rm PC} |$. Linear coupling (LC) models correspond to $q=1$ in Eq. \ref{EQ-BF}. In each figure, the lower panel is a zoom-in of the upper panel, showing with more clarity the corresponding weakly coupling regime. {\it Upper left panel:} LC curves, using the ``latest model'' (labeled as ``IC'' in Figs. \ref{TYPES} and \ref{Jerk_VS-LCDM-COMB-prep}) as a reference for the calculations. {\it Upper right panel:} LC curves, for the type-(i) model (c.f. Fig. \ref{TYPES}). {\it Lower left panel:} PC curves (quadratic coupling, $q=2$) for the ``IC model''. {\it Lower left panel:} PC curves (quadratic coupling, $q=2$) for the ``IC model''.  {\it Lower right panel:} PC curves ($q=6$) for the ``IC model''.\label{Da}}
\end{figure}
\twocolumngrid

\subsection{Constraints for variations to the fine-structure constant \label{DADZ-SEC}}

In this section, we show the stochastic quintessence prediction curves for possible variations in the fine-structure constant, $\Delta \alpha/\alpha (z)$, in the nearby Universe ($z<3$), according to the various coupling forms discussed in Sec. \ref{COUP-SEC}. Fig. \ref{Da} presents the predictions for polynomial couplings (PC), with curves parametrized by the coupling factor $|\zeta | \equiv |\zeta_{\rm PC} |$. The linear coupling (LC, upper panels) models correspond to $q=1$ (c.f. Eq. \ref{EQ-BF}). The curves in Fig. \ref{Da} were obtained directly from Eq. \ref{DA-FORM}, using the results of $\Omega_q(z)$ and $w_q(z)$ for these models, parameterized by $|\zeta | $, spanning $\sim 3$ orders in magnitude. The observational data on $\Delta \alpha(z)/\alpha$ are based on cosmological data (type Ia supernova and Hubble parameter measurements, compiled in \cite{Mar15}) and astrophysical data (high-resolution spectroscopy of QSOs \cite{Mur03}).

For the LC models, we compared the predictions for the ``latest model'' (labeled as ``IC'' in Figs. \ref{TYPES} and \ref{Jerk_VS-LCDM-COMB-prep}) and the type-(i) model (c.f. Fig. \ref{TYPES}). We chose these reference models due to their very distinct behaviors in low redshifts. The IC model (which is an ``extreme'' type-(iii) model) is the ``latest'' viable solution generated from the ensembles, therefore representing the best generator of potential observational signatures at low redshifts. The type-(i) model, on the other hand, presents a peculiar trajectory in phase-space, as noted previously. The type-(ii) models have an intermediary behavior in that redshift range and are not here discussed. For all other predictions, we have fixed the calculations to the ``IC'' model only, given that its curves show steeper behavior for the linear coupling case.

The resulting curves have a ``trumpet''-like form (considering negative and positive values of $\zeta$ as the lower and upper limits, respectively) and are smooth, given that $\Omega_q(z)$ and $w_q(z)$ are smooth in this range (see Fig. \ref{W-COMB}). The ``trumpet''-like form of these models should be compared with the ones analysed in Ref. \cite{Mar15b}: the former have a ``convex''-like form (tending to flatten for $z \gtrsim 2.5$), whereas the later, generally a ``concave''-like one (for fixed $w_q$).  From Eq. \ref{DA-FORM}, the variability in $\alpha$ is directly proportional to the integrated contributions of $\Omega_q$ and $w_q$, with the proportionality factor given by the coupling $\zeta$. In the type-(i) model, the integral is very small, since $w_q(z) \sim -1$ for this model, in that redshift range. This gives a weak dependence of $\Delta \alpha/\alpha$ with redshift, even for relatively high couplings. For the IC model, on the other hand, this weak dependence requires lower coupling values. Yet, it is interesting to note that the equation of state parameter for the IC model varies significantly in that redshift range (i.e., from  $w_q=0$ at $z\sim 3$ to $w_q=-1$ at $z=0$; c.f. Fig. \ref{TYPES}), but the impact of such a large variation is only established for higher coupling values. 

For the polynomial cases (lower panels of Fig. \ref{Da}), the quadratic ($q=2$) and higher ($q=6$) powers clearly have the effect of changing the shape of ``trumpet''-like form of the curves, indicating an earlier and steeper onset of the evolution in $\Delta \alpha/\alpha (z)$. The curves can still accommodate the data with weak enough couplings  $|\zeta | \lesssim 1 \times 10^{-4}$.

\begin{figure} [tbhp]
\includegraphics[width=1\columnwidth]{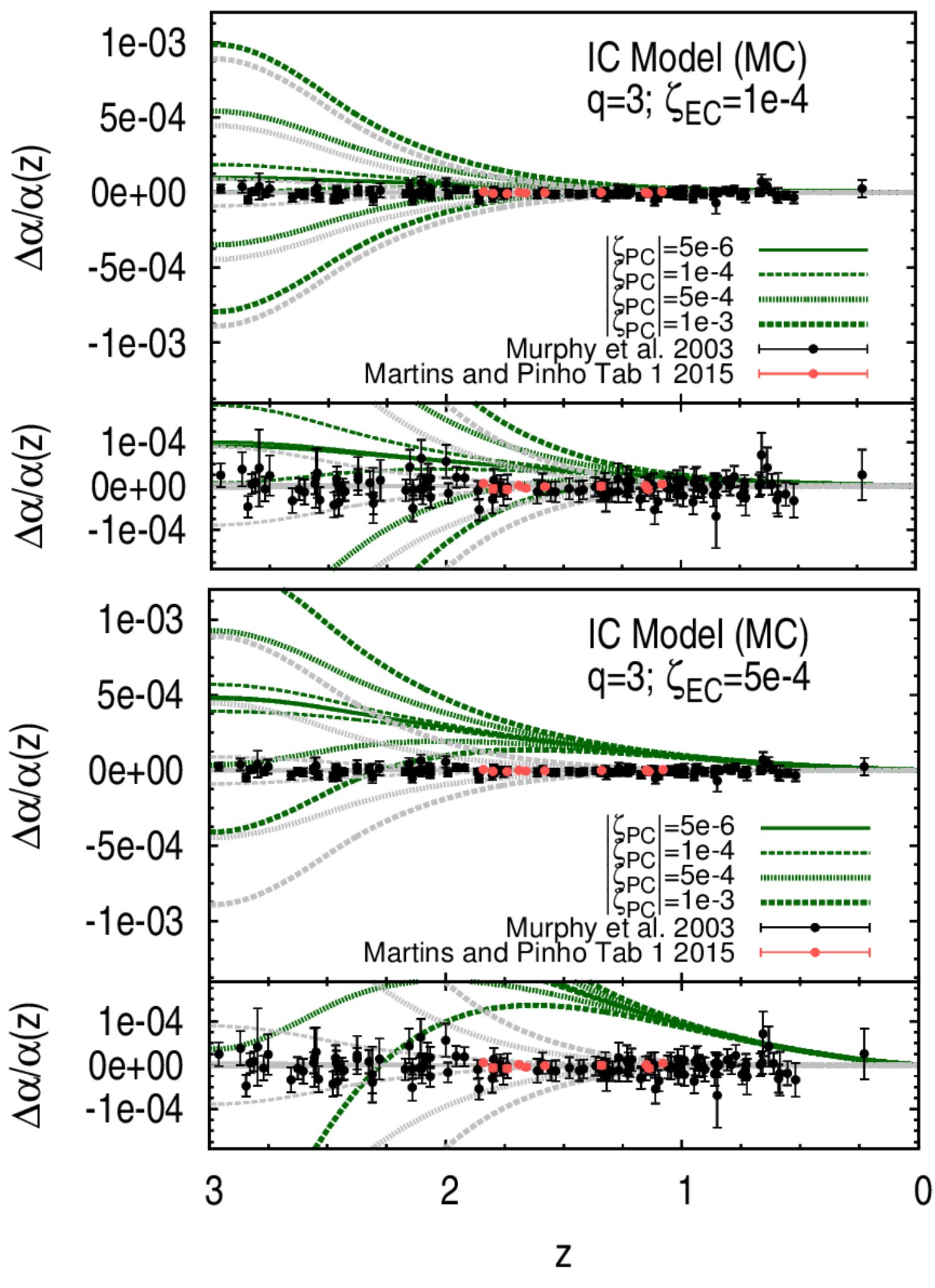}
\caption{(Color online). Stochastic quintessence predictions for the variation $\Delta \alpha/\alpha (z)$ for mixed couplings (MC), adopting $q=3$, and using the ``latest model'' (labeled as ``IC'' in Figs. \ref{TYPES} and \ref{Jerk_VS-LCDM-COMB-prep}) as a reference for the calculations. Curves are parametrized by the coupling factors $|\zeta_{\rm PC} |$ and $\zeta_{\rm EC} $ . In each figure, the lower panel is a zoom-in of the upper panel, showing with more clarity the corresponding weakly coupling regime. Lighter grey curves in all panels correspond to the PC curves with same coupling factors $|\zeta_{\rm PC} |$ but without the exponential term.  {\it Upper panel:} Small deviations from the purely PC curves, for $\zeta_{\rm EC} =  1 \times 10^{-4}$. {\it Lower panel:} Stronger deviations from the purely PC curves, for $\zeta_{\rm EC} =  5 \times 10^{-4}$.\label{Da2}}
\end{figure}

Mixed coupling (MC) cases are shown in Fig. \ref{Da2}. For this analysis, we have adopted $q=3$, and used the ``latest model'' (labeled as ``IC'' in Figs. \ref{TYPES} and \ref{Jerk_VS-LCDM-COMB-prep}) as a reference for the calculations. In order to facilitate the comparison with purely PC counterparts (that is, those with same coupling factors $|\zeta_{\rm PC} |$, but without the exponential term), we have included the results for these cases as well, using lighter grey curves in all panels of Fig. \ref{Da2}. The results show a general trend in the sense that, the higher the values of $\zeta_{\rm EC}$ (say, above $\zeta_{\rm EC} \gtrsim 1 \times 10^{-4}$), the stronger the deviation from the purely PC curves, ultimately leading to a conflict with the experimental data.

\section{Conclusion \label{CONC-SEC}}

We found that stochastically generated, dynamical models of dark energy, discussed by Chongchitnan and Efstathiou (2007), which behave as quintessence models in high redshifts but shift to a cosmological constant behavior at low redshifts, admit solutions with peculiar properties for the observables $j(z)$ and $\Delta \alpha(z)/\alpha$ in low redshifts. Despite of the dynamical richness of the stocastic solutions, they locally (in a ``time sliced'' sense) follow the purely exponential potential models and their behavior can be inferred from known attractor dynamics \cite{Cop98}. In this work, we presented a simple classification of trajectories in phase-space and qualitatively analysed their behavior in terms of the observables $j(z)$ and $\Delta \alpha(z)/\alpha$.

Using a slow-roll approximation to express the jerk parameter as a function of the field variable, we showed that  $j(z)$ deviates (fluctuates) from a pure $\Lambda$CDM model up to a $\sim 5\%$ level in the nearby Universe. However, considering the analysis given by Bochner et al. (2013), where the jerk parameter can only be constrained currently to the large interval $j \sim [-9.2, 9.8]$ in the nearby Universe, potential imprints of the fluctuations found in the present models are, at this stage, unlikely to be detected and must await future data (e.g. \cite{Abb05}).

By assuming a linear coupling between the dark energy and the electromagnetic sectors, we showed that stochastic quintessence models are fully compatible with current observational limits ($z \lesssim 3$) on $\Delta \alpha(z)/\alpha$ and on $\zeta$ (note that constraints are much more stringent, e.g.,  \cite{Mar15}, \cite{Mar15b}). These models offer distinctive observational imprints for $\Delta \alpha(z)/\alpha$ only at high couplings.  The IC model fits the data for any coupling of the order $|\zeta| \lesssim 10 ^{-4}$, and the type-(i) model, for $|\zeta| \lesssim 5 \times 10 ^{-4}$ (and they represent two opposite ``extremes'' of all solutions from the ensembles obtained in this work). However, these models illustrate that variabilities in $\alpha$ are weakly dependent on redshift, for couplings of the order $|\zeta| \sim 10 ^{-4}$, even for large variations in the equation of state parameter at relatively low redshifts. 

For the case of nonlinear couplings, the shape of ``trumpet''-like form of the curves is changed relatively to the linear case, indicating an earlier and steeper onset of the evolution in $\Delta \alpha/\alpha (z)$. In particular, the mixed coupling cases show that, the higher the values of $\zeta_{\rm EC}$ ($\gtrsim 1 \times 10^{-4}$), the stronger the deviation from the purely polynomial curves, ultimately leading to a conflict with the experimental data. All curves can still accommodate the data with weak enough couplings.

Further systematic analysis of viable stochastic quintessence models, exploring the parameter space into more ``extreme'' choices ($z_{\rm trans}$, $\lambda(z)$ amplitudes, etc) are left for future work and would be very interesting to be tested against a larger volume of high-precision cosmological data, e.g. Ref. \cite{Sar15}.

\begin{acknowledgments}
We thank the anonymous referee for valuable suggestions and comments. A.L.B.R. thanks the support of CNPq, under grant 309255/2013-9, and the financial support from the project  Casadinho PROCAD – CNPq/CAPES number 552236/2011-0. 
\end{acknowledgments}


\end{document}